\documentclass[aOAps,superscriptaddress,prl,twocolumn,footinbib]{revtex4-1}
\usepackage[dvipdfmx]{graphicx}
\usepackage{amsmath}
\usepackage{amssymb}
\usepackage{bm}
\usepackage{color}
\usepackage{verbatim}

\begin{document}

\title{
Machine learning of mirror skin effects in the presence of disorder
}
\author{Hiromu Araki}
\email{araki@rhodia.ph.tsukuba.ac.jp}
\author{Tsuneya Yoshida}
\author{Yasuhiro Hatsugai}
\affiliation{Department of Physics, University of Tsukuba, Tsukuba, Ibaraki 305-8571, Japan}
\begin{abstract}
  Non-Hermitian systems with mirror symmetry may exhibit mirror skin effect which is the extreme sensitivity of the spectrum and eigenstates on the boundary condition due to the non-Hermitian topology protected by mirror symmetry.
  In this paper, we report that the mirror skin effect survives even against disorder which breaks the mirror symmetry.
  Specifically, we demonstrate the robustness of the skin effect by employing the neural network which systematically predicts the presence/absence of the skin modes, a large number of localized states around the edge.
  The trained neural network detects skin effects in high accuracy, which allows us to obtain the phase diagram.
  We also calculate the probability by the neural network for each of states.
  The above results are also confirmed by calculating the inversed participation ratio.
\end{abstract}

\maketitle

{\it Introduction.---}Topological properties of condensed matter systems have been extensively studied in these decades
~\cite{RevModPhys.83.1057,PhysRevLett.95.146802,Ryu_2010,PhysRevLett.71.3697,RevModPhys.89.041004}.
Among them, non-Hermitian topology attracts much attention in these years~\cite{PhysRevLett.121.213902,PhysRevLett.116.133903,PhysRevLett.120.146402, PhysRevX.8.031079,Lieu_1DnHTopo_PRB18,PhysRevB.99.075130, PhysRevA.95.062118, PhysRevB.98.094307, PhysRevLett.123.090603,nphys1073, PhysRevLett.115.200402, PhysRevB.100.054301, PhysRevLett.122.237601, PhysRevB.102.024205,PhysRevB.102.041119, PhysRevB.101.020201,PhysRevResearch.2.033052, PhysRevA.101.063612, Liu_2020,PhysRevLett.77.570, PhysRevB.56.8651, PhysRevB.101.235150, PhysRevB.102.035153, PhysRevB.101.121109,s41598-019-53253-8, PhysRevLett.121.086803, PhysRevLett.121.136802, PhysRevLett.123.246801,PhysRevLett.121.026808,Yokomizo_BBC_PRL19,Yokomizo_NBlochBBCEP_arXiv20,PhysRevB.99.081103, s41467-018-08254-y, PhysRevLett.122.076801, PhysRevLett.123.016805,PhysRevLett.123.123601, PhysRevLett.118.040401, PhysRevB.102.054204,s11433-020-1521-9,Bergholtz_Review19,Yoshida_nHReview_PTEP20,Ashida_nHReview_arXiv20}.
The platforms of non-Hermitian topological physics extends to a wide variety of systems such as,
open quantum systems~\cite{PhysRevA.95.062118,PhysRevB.98.094307,PhysRevLett.123.090603,nphys1073,s41598-019-53253-8,Yoshida_nHFQHJ_PRR20},
electric circuits~\cite{PhysRevResearch.2.022062,PhysRevB.100.054301,PhysRevB.101.020201,s41567-020-0922-9,PhysRevResearch.2.033052},
photonic crystals~\cite{PhysRevLett.115.200402,PhysRevLett.121.213902,PhysRevLett.122.237601,PhysRevB.102.024205,Zhou1009,s41467-018-03822-8,Zhao1163,PhysRevLett.115.040402,s41566-019-0453-z},
equilibrium system of quasiparticles~\cite{VKozii_nH_arXiv17,Zyuzin_nHEP_PRB18,Papaji_nHEP_PRB19,Yoshida_EP_DMFT_PRB18,Matsushita_ER_PRB19,Kimura_SPERs_PRB19,Michishita_EP_PRB20,Yoshida_nHReview_PTEP20,Rausch_EP1D_arXiv20,Matsushita_nHResp_arXiv20},
and so on.

In non-Hermitian topological systems, various novel phenomena have been reported which do not have Hermitian counterparts~\cite{Rotter_2009, PhysRevB.99.041406, PhysRevB.99.041202, Zhou:19, PhysRevB.99.121101, PhysRevLett.123.066405, PhysRevB.100.054109, PhysRevA.98.042114,PhysRevLett.121.026808, PhysRevB.99.075130, PhysRevLett.124.056802}.
One of the typical examples is the skin effect~\cite{PhysRevLett.116.133903, PhysRevLett.121.026808, PhysRevLett.121.086803, PhysRevLett.121.136802, PhysRevB.97.121401, PhysRevLett.123.170401, PhysRevLett.123.246801, PhysRevB.99.201103, Zhang_BECskin19,PhysRevLett.124.086801} which can be observed for Hatano-Nelson model~\cite{PhysRevLett.77.570, PhysRevB.56.8651}, a one-dimensional tight-binding model with the non-reciprocal hopping.
This non-reciprocity results in the non-Hermitian topological properties characterized by the winding number~\cite{PhysRevB.99.201103,Zhang_BECskin19,PhysRevLett.124.086801}; the energy spectrum of the Bloch Hamiltonian winds around the origin of the complex plane with increasing the momentum from $-\pi$ to $\pi$.
Corresponding to the non-trivial properties, the system shows extreme sensitivity of the energy spectrum and the eigenstates.
In particular, the non-reciprocity induces skin modes which are a large number of localized states around the edge due to the finite winding number in the bulk.
This remarkable topological phenomenon in non-Hermitian systems is further extended by taking into account symmetry~\cite{PhysRevLett.124.086801,PhysRevResearch.2.022062,Okugawa_HOSkin_arXiv2020,Kawabata_HOSkin_arXiv2020,Fu_HOSkin_arXiv2020}.
In particular, it turned out that mirror symmetry protects a type of non-Hermitian crystalline symmetry inducing the mirror skin effect~\cite{PhysRevB.97.205110}.

Despite the above progress of non-Hermitian crystalline topology, the robustness of the perturbations breaking the relevant symmetry remains unclear. Such perturbations is inevitable in experiments because crystalline symmetry is generically broken by impurities~\cite{Hsieh_SnTe_NatComm2012}.

In this paper, we show that the mirror skin effect survives even in the presence of disorders, which demonstrates that the non-Hermitian crystalline topology is robust against perturbation breaking the crystalline symmetry.
The robustness of the mirror skin effect is elucidated by a machine learning approach~\cite{PhysRevD.88.062003, s41567-019-0554-0,PhysRevB.94.195105, nphys4035, PhysRevE.96.022140, PhysRevLett.118.216401, PhysRevB.96.195145, PhysRevB.99.104106, PhysRevB.99.121104, nphys4037, PhysRevB.97.115453, PhysRevLett.120.066401, PhysRevB.98.085402, PhysRevB.97.134109, PhysRevB.102.054512, s41524-019-0224-x,s41567-019-0512-x, PhysRevB.102.054107, doi:10.7566/JPSJ.89.022001,zhang2020machine, narayan2020machine}.
Namely, we train the neural network in the clean limit so that it predicts the presence/absence of the skin modes in the presence of disorder, which is a signal of the mirror skin effect.
The trained neural network systematically predicts the presence/absence of the skin effects, which allows us to obtain the phase diagram. The obtained phase diagram elucidate the robustness of non-Hermitian topology with mirror symmetry against disorders. We also confirm the robustness by computing the inverse participation ratio (IPR).
Finally, we apply the neural network to each of states and
calculate the probability of the localized states.

The rest of this paper is organized as follows.
Firstly, a toy model with disorder is introduced.
Secondly, effects of disorder are discussed for specific strength of disorder.
The distribution of complex energies and the edge states from the skin effects are also shown.
In next, the model for the machine learning and the method for
training are explained. Then the results by applying the machine learning model
to the non-Hermitian model are shown.
Then, the machine learning model is applied to each of states and
we calculate the probability of the localized states about them.
Lastly, we conclude our study.

{\it Non-Hermitian model.---}In order to demonstrate the robustness of the mirror skin effect against perturbations breaking the relevant symmetry, we analyze a fermionic bilayer system with disorder.
The Hamiltonian reads,
\begin{eqnarray}
\label{eq: Hw}
 H_{\rm W} &=& H_0 + \sum_{j\alpha} w_{j\alpha} c_{j\alpha}^\dagger  c_{j\alpha},
\end{eqnarray}
where $W$ is the strength of the randomness and $w_{j\alpha}$ is a uniform random variable in $[-W/2, W/2]$.
The operator $c^\dagger_{j\alpha}$ ($c_{j\alpha}$) creates (annihilates) a spinless fermion at site $j$ of layer $\alpha$.
In the clean limit $W=0$~\cite{PhysRevResearch.2.022062}, the system is described by $H_0$;
\begin{eqnarray}
\label{eq: H0}
 H_0 &=& \sum_{\bm{k}\alpha\beta} c^\dagger_{\bm{k}\alpha} h_{\alpha\beta}(\bm{k}) c_{\bm{k}\alpha}, \\
 h(\bm{k}) &=& [2t(\cos k_x + \cos k_y)-\mu] \rho_0 \nonumber \\
 &&+ i \Delta \sin k_x \rho_3 + i \Delta \sin k_y \rho_2,
\end{eqnarray}
where $c^\dagger_{j\alpha}$ ($c_{j\alpha}$) denotes the Fourier transformed creation (annihilation) operator. The vector $\bm{k}=(k_x,k_y)$ $( -\pi \leq k_{x(y)} < \pi)$ describes the momentum of the two-dimensional system.
The matrices $\rho_i$ ($i=0,1,2,3$) are the Pauli matrices acting on the pseudo-spin space of the two layers.
The Hamiltonian in the clean limit $H_0$ preserves the mirror symmetry;
$M_x H(\bm{k})M_x^{-1} = H (M_x \bm{k})$
with
$M_x = \rho_2 P_x$.
Here, $P_x$ flips the momentum from $\bm{k} = (k_x, k_y)$ to $M_x\bm{k} := (-k_x, k_y)$.

The mirror symmetry of $H_0$ protects the non-Hermitian point-gap topology~\cite{point-gap_ftnt} which is characterized by the mirror winding number~\cite{PhysRevResearch.2.022062};
\begin{equation}
  \nu_M (k_x^*) = \sum_{m=\pm} \mathrm{sgn}(m) \int \frac{dk_y}{4\pi i} \partial_{k_y} \log \det [h_m (k_x^*, k_y) - E_{\rm pg}],
\end{equation}
with $\mathrm{sgn}(m)$ taking $1$ ($-1$) for $m=+$ ($m=-$).
Here, $h_{+(-)}$ is the Bloch Hamiltonian for plus (minus) sector of the mirror operator $M_x$~\cite{M_and_h_ftnt}.
For $k^*_x=0$, the mirror winding number takes $\nu_M (0)=-1$ with $E_{\rm pg} = 2t - \mu$, while for $k^*_x=\pi$, it takes $\nu_M (\pi)=-1$ with $E_{\rm pg} = -2t - \mu$.

{\it Effect of disorder on the mirror skin effect.---}Prior to the systematic analysis based on the machine learning approach, we discuss the effect of disorder for specific values of disorder strength, $W=1$ and $W=10$.

\begin{figure}[t]
  \begin{center}
    \includegraphics[width=\linewidth]{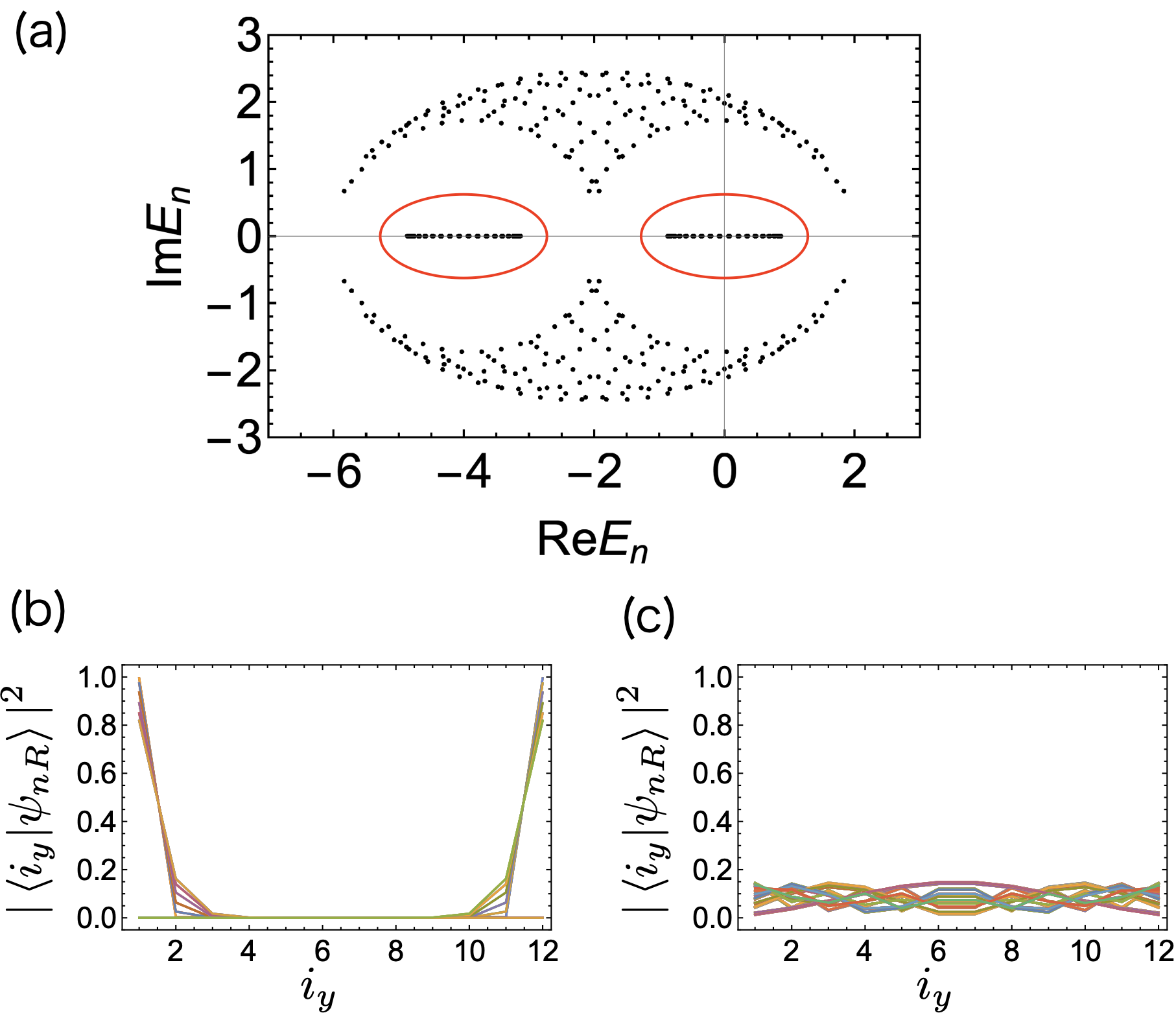}
    \caption{
    (a): The eigenenergies of the model with ($t$, $\mu$, $\Delta$) = ($1$, $2$, $1.8$)
    and $M=18$. The boundary conditions are periodic for the $x$-axis and open for the $y$-axis.
    (b): The density plot of the eigen states corresponding to the eigenenergies in
    the red circles in panel (a). The particles are localized on edges.
    (c): The density plot of the eigen states corresponding to the eigenenergies outside
    the red circles in panel (a). The particles are itinerant in the bulk.
    Panels (b) and (c) are obtained for $M=12$.
    }
    \label{fig:1}
  \end{center}
\end{figure}

For comparison, we firstly analyze the spectrum and eigenstates in the clean limit.

Figure~\ref{fig:1} shows the spectrum and eigenstates for the system of $M\times M$ sites. Here, the boundary conditions are periodic for the $x$-axis and open for the $y$-axis.
[Fig.~\ref{fig:1}(a) is obtained for $M=18$ while Fig.~\ref{fig:1}(b)~and~\ref{fig:1}(c) are obtained for $M=12$].

Figure~\ref{fig:1}(b) indicates that due to the topology characterized by the mirror winding number, the $4M$-states enclosed in red circles in Fig.~\ref{fig:1}(a) are localized. These skin modes are signals of the skin effects.
Here, the number $4M$ can be calculated as follows:
at given $k^*_x$ taking $0$ or $\pi$, the mirror skin effect induces $2M$-localized states, where the prefactor $2$ is arising from the number of layers.

We note that the all of states outside the red circles are extended to the bulk [see Fig. ~\ref{fig:1}(c)].

\begin{figure}[t]
  \begin{center}
    \includegraphics[width=\linewidth]{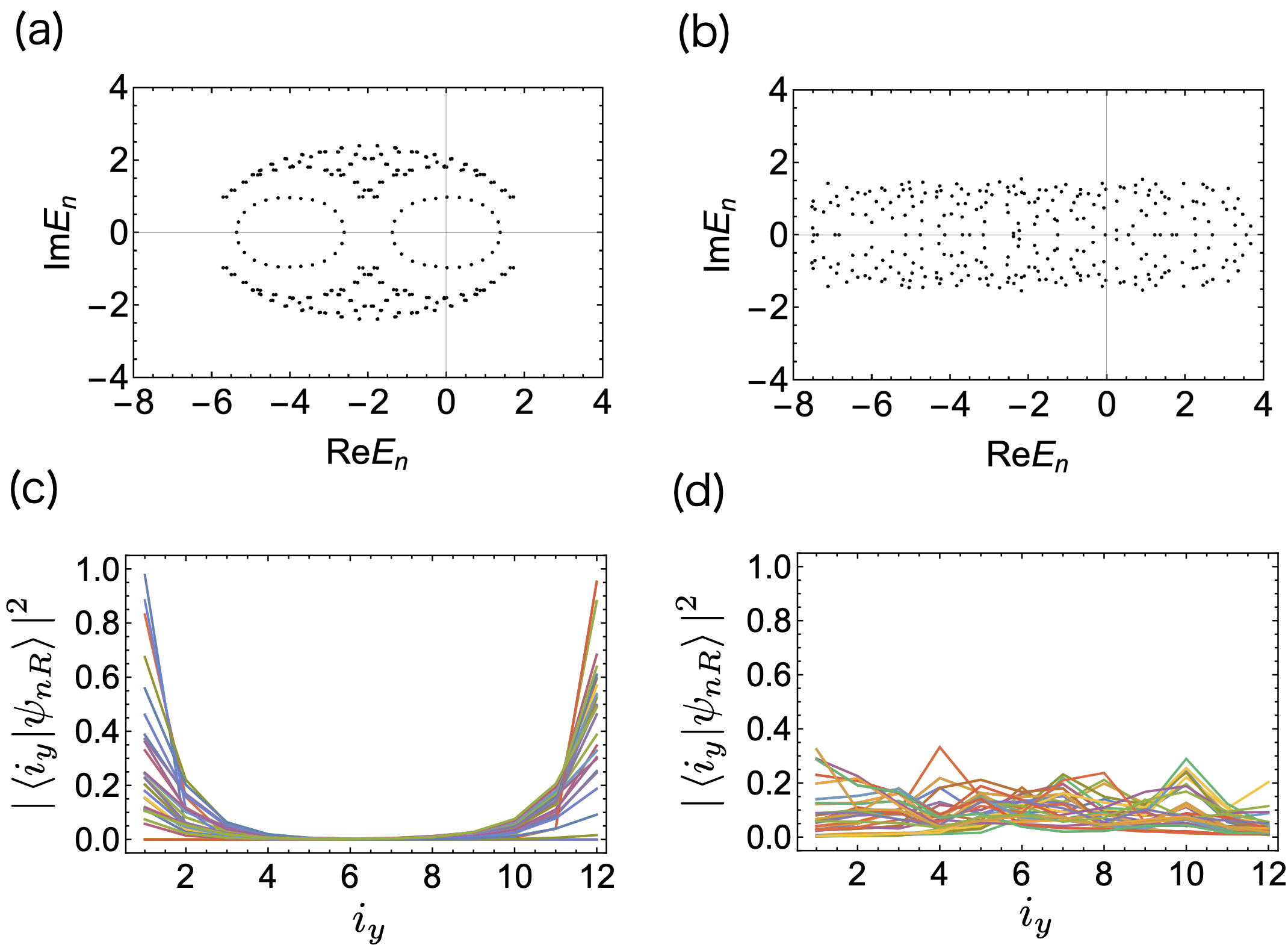}
    \caption{
(a) and (b): The eigenenergies of the model with ($t$, $\mu$, $\Delta$) = ($1$, $2$, $1.8$) and $M=18$.
(b) and (d): The density plot of the $4M$ eigenstates around $E^+_{\rm pg}$ and $E^-_{\rm pg}$ for $M=12$. Panels (a) and (c) [(b) and (d)] are obtained for $W=1$ [$W=10$].
    }
    \label{fig:2}
  \end{center}
\end{figure}

\begin{figure*}[tb]
  \begin{center}
    \includegraphics[width=0.9\textwidth]{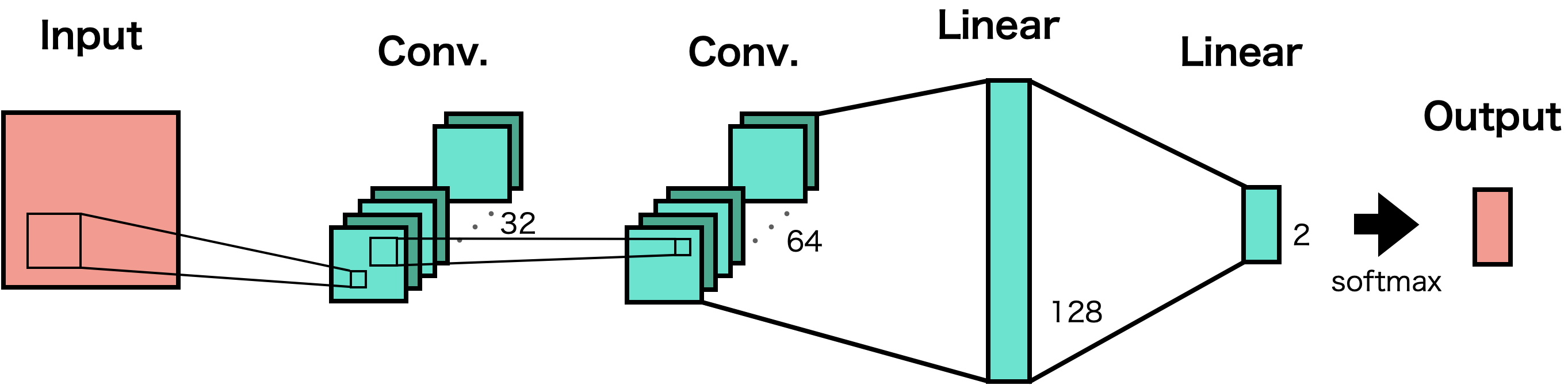}
    \caption{
    Schematic picture of our CNN model.
    Input data are batched two dimensional images.
    The model consists of two
    CNN layers and two linear layers. The numbers in this figure
    are the output layers for the CNN layers and the output parameters
    for the linear layers.
    Calculating the softmax for the output of the second linear layer, the output data
    of the model are obtained.
    }
    \label{fig:3}
  \end{center}
\end{figure*}

We analyze the system for specific values of disorder strength $W=1,10$.
The obtained results imply that the point-gap topology remains for weak $W$, although introducing the disorder breaks the mirror symmetry.
Further increasing the interaction suppresses the skin effects as the localized states are buried in the bulk states.

Figure \ref{fig:2} shows the energies and a part of eigenstates of the model with randomness.
Boundary conditions are the same as the Fig.~\ref{fig:1} --  periodic for the $x$-axis and
open for the $y$-axis.
Figures \ref{fig:2}(a) and \ref{fig:2}(b) show the eigenenergies of the model with
($t$, $\mu$, $\Delta$, $W$) = ($1$, $2$, $1.8$, $1$) and
($t$, $\mu$, $\Delta$, $W$) = ($1$, $2$, $1.8$, $10$), respectively.
The figures show that the edge-localized states from the skin effect are still
in the gap of $E_{\rm pg}^\pm$ for $W=1$, however the gap is closed for $W=10$.
Next we consider the eigenstates around the point $E_{\rm pg}^\pm$.
Figures \ref{fig:2}(c) and \ref{fig:2}(d) show the density plot of the $4M$ states around $E^\pm_{\rm pg}$ for $W=1$ and $W=10$, respectively.
For $W=1$, the in-gap states are still localized at edges. It suggests that the skin effect still remains.
However, for $W=10$, the gap is closed and there are no localized states around the $E^\pm_{\rm pg}$.

The above results imply that the mirror skin effect survives even in the presence of the disorder. In next, we elucidate the robustness by systematic analysis based on the machine learning technique.

{\it Machine learning of disordered non-Hermitian systems.---}Hereafter, we try to draw the phase diagram of the non-Hermitian system with randomess.
To accomplish it, we use a machine learning.
By using the convolutional neural network (CNN), we can detect the edge states and draw a phase diagram systematically for
each parameters~\cite{PhysRevB.97.205110, PhysRevB.99.085406}.

As input data, we take the absolute square of the wave function.
The boundary conditions of the system are periodic for
the $x$-axis and open for the $y$-axis and system size is $2M^2$.
Here after, we set $M=6$.
Since the system is in two dimensions and has two degrees of freedom at each point,
each of input data is $(M \times M \times 2)$ matrix.
In addition, we take average of the $4M$ wave functions around the reference
energies $E^\pm_{\rm pg}$.

The output data are whether the edges states from the skin effect
are in the point-gap of the $E^\pm_{\rm pg}$.
We use the data of the clean systems as the training data.
Then, the trained model predicts the phase of the disordered systems.


\begin{figure}[thb]
  \begin{center}
    \includegraphics[width=\linewidth]{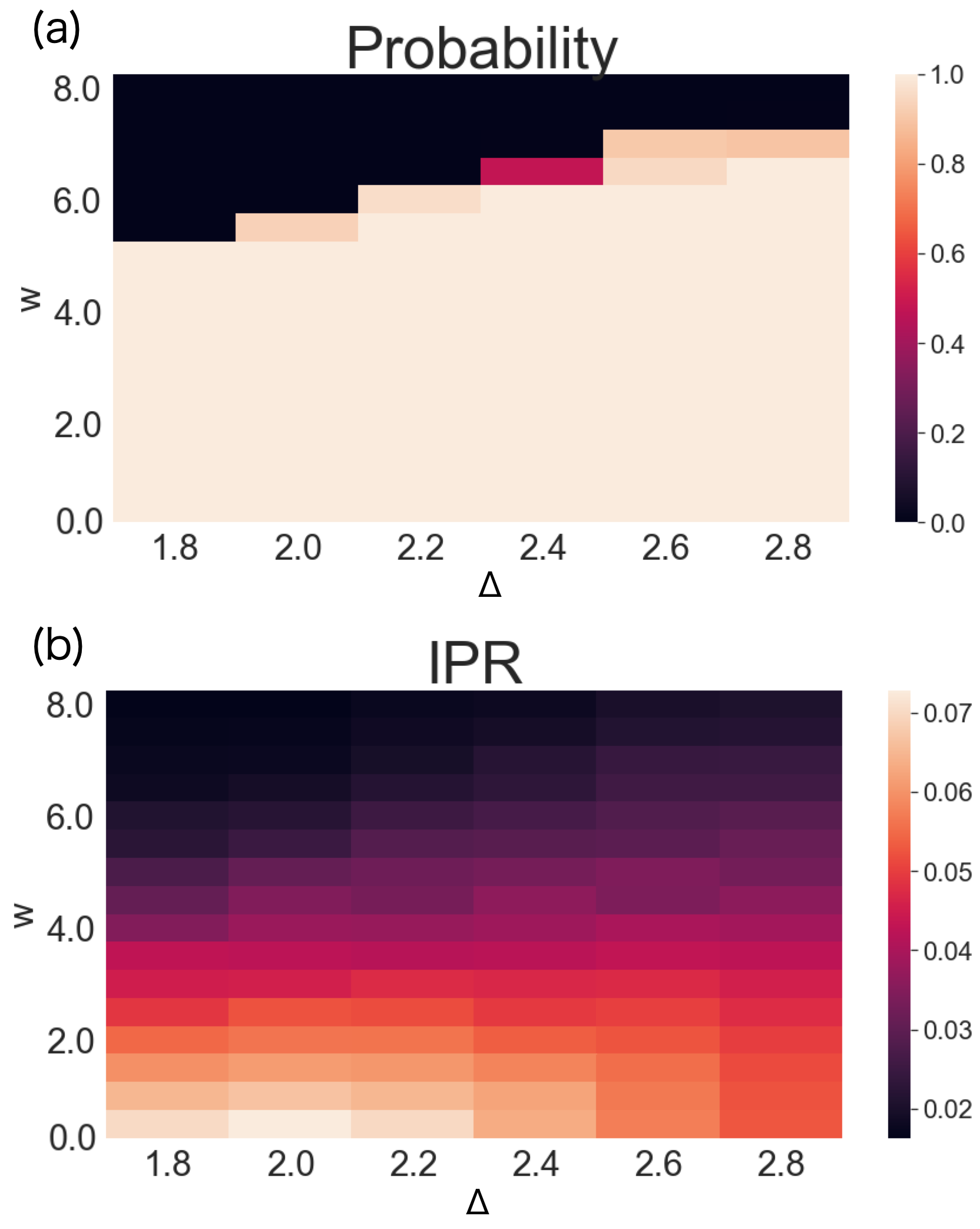}
    \caption{
    (a) Probabilities that are culculated by the machine learning model for
    the input data specificated by $W$ and $\Delta$.
    (b) The IPR for the input data specificated by $W$ and $\Delta$.
    }
    \label{fig:4}
  \end{center}
\end{figure}

Here after, we show the architecture of the neural network.
The schematic picture of our model is shown in Fig.~\ref{fig:3}.
It consists of two convolutional layers and two linear layers.
The number of filters comprising the convolutional layers is 32 and 64.
The kernel size of each convolutional layer is $3\times 3$.
The first linear layer transforms the output of the previous layer to
128 dimensional vector and the second one transforms it to two-dimensional
vector.
The output data are obtained by applying softmax functions
$y_i = e^{x_i} / \sum_{k=1}^2 e^{x_k}$, where $i$ labels one of the phases.
All the activation functions are the ReLU functions.
For input $x$, the ReLU function returns $x$ if $x > 0$, otherwise it returns $0$.
To prevent overfitting, the dropout layer is
inserted after the second convolutional layer and the first linear layer.
The probability of dropout is set to $0.5$.

\begin{figure*}[thb]
  \begin{center}
    \includegraphics[width=\linewidth]{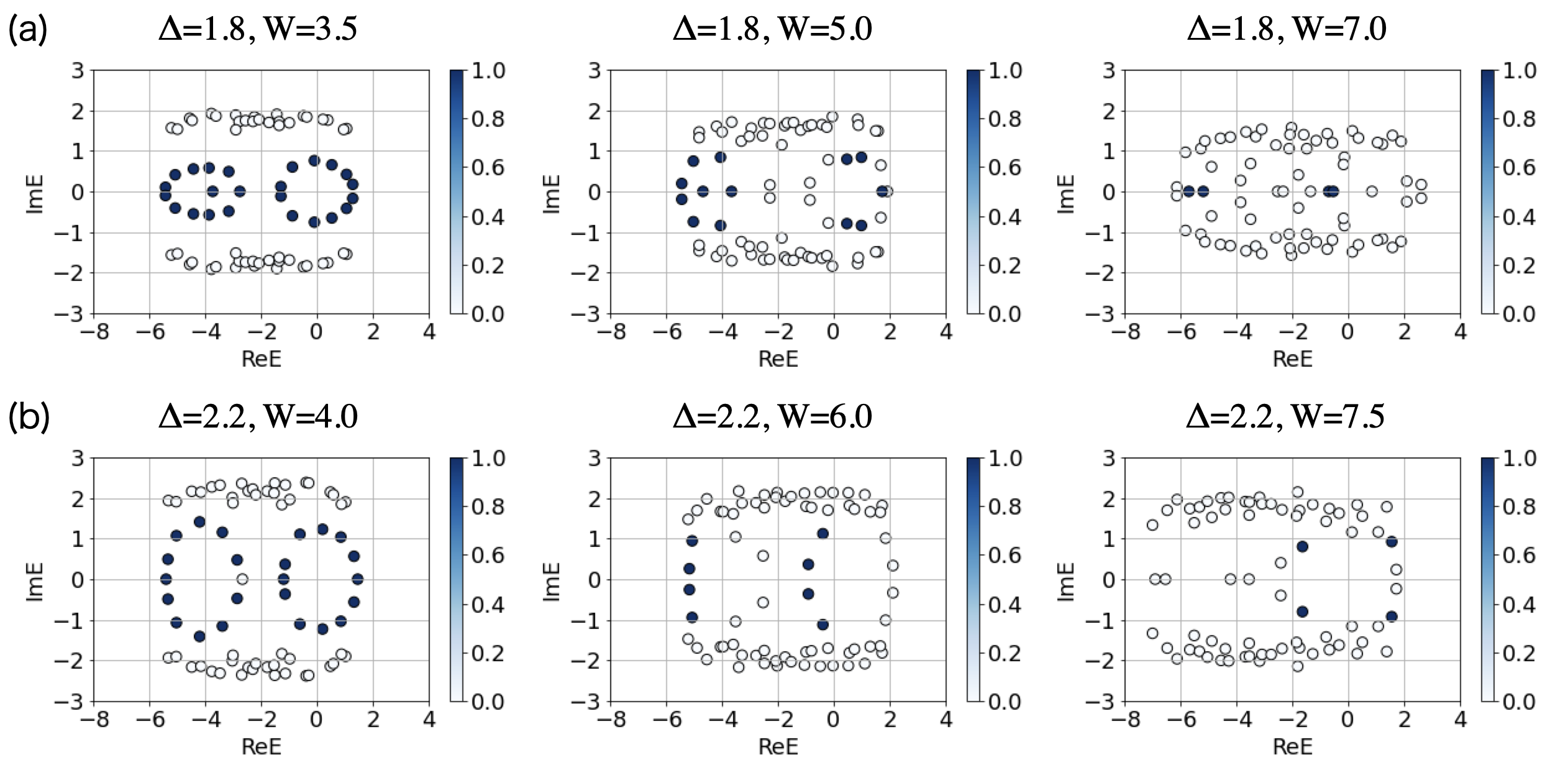}
    \caption{
    The energies of the system with various $\Delta$ and $W$.
    (a) $\Delta$ is fixed to $1.8$.
    (b) $\Delta$ is fixed to $2.2$.
    The color on the dots corresponds with the probability of the localized states
    calculated by the machine learning model.
    }
    \label{fig:5}
  \end{center}
\end{figure*}

To train the model, we use the states of the system in the clean limit.
As stated before, the localized states are in the two distinct point-gaps.
In the following, we explain about the input data in training process.
Firstly, we take the in-gap $4M$ edge states.
Then, we take the average of the probability density of the $4M$ states as the data of localized states.
On the other hand, there are $2M^2 -4M$ bulk states.
In a similar way, we take average of the probability density of the $2M^2 -4M$ bulk states as the
data of bulk states.
We take 2000 samples as input data by randomly selecting $\Delta \in (1.8, 2.8)$.

The output of the model is a two dimensional vector that is interpreted as the probability
of the localized state and the bulk states.
We use 1600 samples for training data and the rest for validation data.
After the training process, the percentage of correct answers is 99.8\% for the validation data.


Here after, we investigate the phase diagram of the disordered non-Hermitian
system.
Figure~\ref{fig:4} shows the phase diagram with $\Delta$ and $w$.
For the disordered system, the input data are prepared as follows.
Firstly, the $2M$ states near the each of reference energies $E^\pm_{\rm pg}$ are selected.
Then the input data are the average of the probability density of the states.
The output data of the trained model tells us whether the states are localized or bulk states.
Figure~\ref{fig:4} (a) shows the probability of the localized states for $\Delta$ and $w$.
The figure shows that the two phases are clearly distinct.
It implies that the skin effect still exists for small $w$ but
the localized states are buried in the bulk states for $w$ lager than
the critical value, which is determined by the competition between the size of point-gaps
and the strength of disorder.
In this model, the size of the point-gap is determined by $t$ and $\Delta$.
Then, if the strength of disorder $w$ exceeds the critical value, the phase change
to the disordered phase (Fig. ~\ref{fig:4}(a)).

Next, we consider IPR as an indicator for the localization.
The IPR $p$ is calculated as
$p = \sum_{k, j} |\phi_k(r_j)|^4$
for a state $\phi$ where $r_j$ is the $j$th lattice point and $k$ is the index of inner degree
of freedom.
Figure~\ref{fig:4} (b) shows the IPR for the averaged states.
The figure shows that the large (small) $p$ corresponds with the localized (bulk) state.
Furthermore, the phase diagram is consistent with the result by the trained machine learning
model.

{\it Machine learning of disordered non-Hermitian systems.---}Here, we apply the machine learning model to the each of states
of the non-Hermitian model with impurities so that
the probability of the localized state is calculated by the machine learning model.
To accomplish it, we use both the averaged states and the non-averaged states as the
supervised data.
This is because the non-averaged states are localized at one of the edges, which are
not appeared in the averaged states.

The supervised data contain 4000 averaged states and 4320 non-averaged
states. Half of the states are the bulk states and the rest states are the localized states.
The system and the architecture of the neural network are the same as before.

Figure ~\ref{fig:5} shows the energies of the system with several values of $\Delta$ and $W$.
The color on the dots corresponds with the probability of the localized states.
In Fig. ~\ref{fig:5}(a) and ~\ref{fig:5}(b), $\Delta$ is fixed to $1.8$ and $2.2$, respectively.
In both Fig. ~\ref{fig:5}(a) and ~\ref{fig:5}(b), the system shown in left, right and center figures
corresponds with the white, dark and boundary region in Fig. ~\ref{fig:3}.
The figures show that there are more localized states in the in-gap states for weak $W$
and there are more extended states for large $W$.
At the phase transition point, both are included.


{\it Conclusion.---}Employing the neural network, we have demonstrated that the mirror skin effect survives even in the presence of the disorder, which elucidates the robustness of the non-Hermitian crystalline topology against perturbations breaking the relevant symmetry.
The neural network, which is trained in the clean limit, systematically predicts the presence/absence of the skin modes in high accuracy, which allows us to obtain the phase diagram. The obtained phase diagram indicates that even in the presence of disorder, the system shows the skin modes due to the point-gap topology with mirror symmetry.
We have also confirmed that the above results are consistent with the analysis based on the IPR.
Finally we have applied the machine learning model to each of states and calculated the
probability of the localized states about them.
The robustness of non-Hermitian crystalline topology against disorders is considered to be experimentally verified by electric circuits.


{\it Acknowledgments.---}We thank Tomonari Mizoguchi for collaboration in Ref.~\onlinecite{PhysRevResearch.2.022062} proposing mirror skin effect.
This work is supported by JSPS Grant-in-Aid for Scientific Research on Innovative Areas ``Discrete Geometric Analysis for Materials Design": Grants No.~JP20H04627.
This work is also supported by JSPS KAKENHI Grants No.~JP17H06138, No.~JP19J12315, and No.~JP19K21032.


\begin{thebibliography}{120}
\bibitem{M_and_h_ftnt}
{ Along the mirror invariant line, specified by $k_x^* = 0$ or $\pi$, the
  Hamiltonian is block-diagonalized about the operator.}
\end{thebibliography}


\end{document}